\documentclass[twoside,11pt]{article}
\usepackage{a4wide}
\usepackage{cite,latexsym,amsfonts,amssymb,exscale,epsfig,bbm}
\usepackage[centertags,sumlimits,intlimits,namelimits,reqno]{amsmath}
\usepackage{amsthm}

\def\href#1#2{#2}%
\def\dhref#1#2{}%

\theoremstyle{definition}

\numberwithin{equation}{section}
\def\address#1{\date{{\sl #1}\\\ \\\theversion}\gdef\date##1{}}%
\def\version#1{\gdef\theversion{#1}}%

\def\dopreprint{\hfill{\small\thepreprint}\\}%
\def\preprint#1{\def\thepreprint{#1}}%
\def\thepreprint#1{}%

\def\pacs#1{\noindent PACS: #1\par}%
\def\keywords#1{\noindent key words: #1\par}%
\def\acknowledgments{\section*{Acknowledgments}}%

\newcommand{\ontop}[2]{\genfrac{}{}{0pt}{2}{\scriptstyle #1}{\scriptstyle #2}}

\def\href#1{}
\def\dhref#1#2{}

\def\nn{\notag}
\def\del{\partial}
\def\emph#1{{\sl #1\/}}

\def\SO{{SO}}

\def\SU{{SU}}
\def\SL{{SL}}
\def\Spin{{Spin}}

\def\C{{\mathbbm C}}
\def\N{{\mathbbm N}}
\def\R{{\mathbbm R}}
\def\Z{{\mathbbm Z}}

\def\ie{{\sl i.e.\/}}

\def\etc{{\sl etc.\/}}
\def\cf{{\sl cf.\/}}
\def\1{\mathbf{1}}

\def\sym#1{{\mathcal #1}}

\let\phi=\varphi
\let\theta=\vartheta
\let\epsilon=\varepsilon

\def\Stab{\mathop{\rm Stab}\nolimits}

\def\mycaption#1#2{%
  \begin{quote}
  \caption{\label{#1}#2}
  \end{quote}}

\let\hat=\widehat
\let\tilde=\widetilde

\makeatletter

\newfont{\@aidxte}{cmsy10}
\newfont{\@aidxel}{cmsy10 scaled 1095}
\newfont{\@aidxtw}{cmsy10 scaled 1200}
\newlength\@aidxtexvi
\newlength\@aidxtexvii
\newlength\@aidxelxvi
\newlength\@aidxelxvii
\newlength\@aidxtwxvi
\newlength\@aidxtwxvii
\newcommand{\alignidx}[1]{%
  \@aidxtexvi=\fontdimen16\@aidxte
  \@aidxtexvii=\fontdimen17\@aidxte
  \@aidxelxvi=\fontdimen16\@aidxel
  \@aidxelxvii=\fontdimen17\@aidxel
  \@aidxtwxvi=\fontdimen16\@aidxtw
  \@aidxtwxvii=\fontdimen17\@aidxtw
    {\mbox{$%
    \fontdimen16\@aidxte=2.9pt
    \fontdimen17\@aidxte=2.9pt
    \fontdimen16\@aidxel=3.1pt
    \fontdimen17\@aidxel=3.1pt
    \fontdimen16\@aidxtw=3.3pt
    \fontdimen17\@aidxtw=3.3pt
    #1$}}%
    \fontdimen16\@aidxte=\@aidxtexvi
    \fontdimen17\@aidxte=\@aidxtexvii
    \fontdimen16\@aidxel=\@aidxelxvi
    \fontdimen17\@aidxel=\@aidxelxvii
    \fontdimen16\@aidxtw=\@aidxtwxvi
    \fontdimen17\@aidxtw=\@aidxtwxvii}

\makeatother

\version{11 December 2002}%
\preprint{}%

\begin{document}
%
% ==============================================================================

\title{\dopreprint On the causal Barrett--Crane model: measure, coupling\\
                   constant, Wick rotation, symmetries and observables}
\author{Hendryk Pfeiffer\thanks{e-mail: hpfeiffer@perimeterinstitute.ca}}
\address{Perimeter Institute for Theoretical Physics,\\
         35 King Street N, Waterloo, Ontario, N2J 2W9, Canada\\
         and\\
         Emmanuel College, St Andrew's Street,\\ 
         Cambridge CB2 3AP, United Kingdom}
\date{\version}
\maketitle

% ==============================================================================
%
\begin{abstract}
%
% ==============================================================================

  We discuss various features and details of two versions of the
  Barrett--Crane spin foam model of quantum gravity, first of the
  $\Spin(4)$-symmetric Riemannian model and second of the
  $\SL(2,\C)$-symmetric Lorentzian version in which all tetrahedra are
  space-like. Recently, Livine and Oriti proposed to introduce a
  causal structure into the Lorentzian Barrett--Crane model from which
  one can construct a path integral that corresponds to the causal
  (Feynman) propagator. We show how to obtain convergent integrals for
  the $10j$-symbols and how a dimensionless constant can be introduced
  into the model. We propose a `Wick rotation' which turns the rapidly
  oscillating complex amplitudes of the Feynman path integral into
  positive real and bounded weights. This construction does not yet
  have the status of a theorem, but it can be used as an alternative
  definition of the propagator and makes the causal model accessible
  by standard numerical simulation algorithms. In addition, we
  identify the local symmetries of the models and show how their
  four-simplex amplitudes can be re-expressed in terms of the ordinary
  relativistic $10j$-symbols. Finally, motivated by possible numerical
  simulations, we express the matrix elements that are defined by the
  model, in terms of the continuous connection variables and determine
  the most general observable in the connection picture. Everything is
  done on a fixed two-complex.
\end{abstract}


% ==============================================================================
%
\section{Introduction}
%
% ==============================================================================

Spin foam models have been proposed as candidates for a quantum theory
of gravity, see, for example, the review articles~\cite{Ba99,Or01a}. A
spin foam~\cite{Ba98a} whose symmetry group is a suitable Lie group
$G$, is an abstract oriented two-complex consisting of faces, edges
and vertices, together with a colouring of the faces with
representations of $G$ and a colouring of the edges with compatible
intertwiners (representation morphisms) of $G$. Spin foam models are
defined by a path integral in terms of a sum over spin foams, often
over all colourings of a fixed two-complex or in addition over a class
of two-complexes.

The most carefully studied model in this context is the Barrett--Crane
model~\cite{BaCr98} which was initially formulated for a Riemannian
signature and a local $\Spin(4)$-symmetry. A version with Lorentzian
signature and $\SL(2,\C)$-symmetry can be constructed along similar
lines. Here we are interested in the model~\cite{BaCr00} in which all
tetrahedra are space-like, \ie\ if the model is formulated on the
two-complex dual to a triangulated four-manifold, then the model
assigns a geometry to the two-complex such that each tetrahedron has a
time-like normal vector.

The idea for the construction of the Barrett--Crane
model~\cite{BaCr98,Ba98a,BaBa00} can be sketched as follows. General
Relativity in four dimensions is reformulated as a topological
$BF$-theory with symmetry group $\Spin(4)$ or $\SL(2,\C)$, depending
on the signature, subject to bi-vector constraints which break the
topological properties and which ensure that the theory is classically
equivalent to General Relativity, possibly allowing degenerate
metrics. Topological $BF$-theory is then regularized and quantized on
a triangulated four-manifold which results in a topological spin foam
model. The bi-vector constraints are finally implemented into this
quantum theory. The result is a spin foam model which dynamically
assigns geometric data to a purely combinatorial triangulation.

The path integral of the spin foam model can then be used in order to
define the matrix elements of some operator between spin network
states. There have been different conjectures, for example, that it
is some unitary `time evolution' operator or that this operator is the
projection from some kinematical Hilbert space onto the physical
Hilbert space of quantum gravity. The precise role of the fourth
direction (`time' ?) in this path integral, however, remained
obscure. In particular it was observed~\cite{BaCh02}, see
also~\cite{Pf02c}, that the amplitudes of this path integral are
positive real so that it does not look like a complex oscillating
`real time' path integral at all. Formally, it could be an Euclidean
(`imaginary time') path integral, but this was not the intention of
the construction, and a physical interpretation of this picture is
also lacking.

Recently, Livine and Oriti~\cite{LiOr02} proposed a modification of
the amplitudes of the Lorentzian Barrett--Crane model in which they
employ, for each pair of triangle and four-simplex, only one out of
two summands of the amplitude with a particular sign in the
exponent. This guarantees that the construction is compatible with a
causal structure imposed on the four-simplices and that the model
resembles a `real time' Feynman (\ie\ causal) path integral of a
Quantum Field Theory with four-dimensional Lorentzian Regge action. In
the following, we call this version of the model the \emph{causal}
Lorentzian Barrett--Crane model. Livine and Oriti~\cite{LiOr02} derive
consistency conditions on the relevant signs for the construction of
this model.

A causal version of the Riemannian Barrett--Crane model can be defined
by analogy. This, however, is not more than just a toy model because
in this case the causal structure has to be imposed completely by hand
and is no longer related to the signature of the metric. We do include
this possibility in the following because it is occasionally very
helpful for technical reasons.

Given the causal model with its definition of the Feynman path
integral, there are a number of natural questions to ask. What is the
status of the measure? Are the integrals convergent, at least for a
fixed triangulation and a fixed assignment of representations to the
triangles? Having identified an action for the path integral, is there
any coupling constant in the model which can affect the dominant
contributions to the path integral? Is there a consistent `Wick
rotation' in order to render all amplitudes positive real and to
obtain an `imaginary time' model whose physical interpretation we
understand? Which numbers can we extract from the model?  In the
present article, we address various aspects of these questions.

In particular, we give an explicit construction of suitable sign
factors that satisfy the conditions of Livine and
Oriti~\cite{LiOr02}. We rephrase the causal model so that it becomes
manifest that a future-pointing increase in the lapse function of the
path integral (for details, see~\cite{LiOr02}) corresponds to
four-simplices with positive four-volume. We demonstrate how to split
measure and $e^{iS}$-amplitude so that all integrals originating from
the $10j$-symbols are well defined, and we introduce a dimensionless
coupling constant into the model.

As far as the Wick rotation is concerned, we follow ideas from the
area of dynamical triangulations~\cite{AmJu01} and proceed
four-simplex by four-simplex, introducing $i$ or $-i$ into the
exponents in order to obtain an Euclidean action, depending on whether
the simplex itself or its opposite oriented counterpart appears in the
Feynman propagator. It should be pointed out that the relation of this
Euclidean theory with the original one does not yet have the status of
a theorem comparable to the situation in axiomatic quantum field
theory. What is the status of Wick rotation in quantum gravity in
general? In a path integral of quantum gravity one has to sum, in one
way or another, over all possible four-metrics most of which do not
admit any global time coordinate which one could use in order to Wick
rotate a Lorentzian to a Riemannian manifold. In addition, the choice
of a time coordinate is obviously not a diffeomorphism invariant
concept. These problems render Wick rotation in a generic quantum
gravity setting highly questionable. Why then discuss Wick rotation at
all?

The answer is that one should not take Wick rotation literally and not
try to substitute $t\mapsto -i\tau$ for some coordinate $t$. Rather,
there exists a generalization~\cite{AsMa00} of the
Osterwalder--Schrader Euclidean reconstruction theorem to background
independent theories which relates a path integral theory which is
defined on a four manifold of the topology $\Sigma\times\R$ and whose
measure is reflection positive, with a canonical quantum theory in
terms of Hilbert space and Hamiltonian on $\Sigma$. In a spin foam
context, such a procedure will be independent of the metric signature
because the local frame symmetry is treated as an internal symmetry in
the underlying first order formulation of General Relativity. We
therefore expect that we can independently euclideanize both the
Riemannian and the Lorentzian signature models and that their
signatures and symmetry groups do not change under such a
transformation.

Since up to now, there does not exist any generally accepted way of
achieving independence of the Barrett--Crane model from the chosen
triangulation, we cannot restore the full diffeomorphism
symmetry. Therefore we cannot yet verify all of the generalized
Osterwalder--Schrader axioms~\cite{AsMa00}.

The main motivation for searching an Euclidean version of the causal
model is the fact that the resulting model with its positive real
amplitudes can be tackled by standard simulation algorithms. We call
the resulting `Wick rotated' model the \emph{Euclidean} (Riemannian or
Lorentzian) Barrett--Crane model. Note that the term \emph{Euclidean}
refers to the use of `imaginary time' as opposed to the term
\emph{Riemannian} which denotes the metric
signature\footnote{Unfortunately, the term Euclidean was historically
also used in order to denote the Riemannian signature.}

In the resulting Euclidean model, one wishes to calculate interesting
quantities. We address the question of which are suitable observables
(here meaning numbers we can extract from the model on a fixed
two-complex) in the \emph{connection picture}, the reformulation which
uses continuous variables and which was developed for the Riemannian
model in~\cite{Pf02a} and calculate the most general function of the
connection variables that is compatible with the local symmetries of
the model. Finally, starting from the analysis of the local
symmetries, we show how both the causal and the Euclidean model can
still be re-expressed in terms of relativistic $10j$-symbols which are
familiar from the original model.

The present article is organized as follows. In
Section~\ref{sect_notation}, we introduce our notation for oriented
two-complexes and introduce the Riemannian and Lorentzian
Barrett--Crane models in their original formulation. In
Section~\ref{sect_measure}, we review the causal versions of these
models, present a construction of all required signs, carefully choose
an appropriate measure and identify a dimensionless coupling
constant. The transition amplitudes are euclideanized in
Section~\ref{sect_wick}. We study in Section~\ref{sect_observe} the
most general observables of the model in the connection picture. In
Section~\ref{sect_back}, we show how the causal and the Euclidean
models can be rephrased in terms of the relativistic $10j$-symbols,
similar to the original formulation of the
model. Section~\ref{sect_conclusion} contains some concluding
comments.

% ==============================================================================
%
\section{Notation and Conventions}
%
% ==============================================================================
\label{sect_notation}

%------------------------------------------------------------------------------
\subsection{Triangulations and two-complexes}
%------------------------------------------------------------------------------
\label{sect_triang}

We consider the triangulation of an oriented piecewise linear
four-manifold $M$ and its dual two-complex. This two-complex is
described by sets $V$ of vertices, $E$ of edges and $F$ of faces
together with maps indicating the source $\del_-(e)\in V$ and target
$\del_+(e)\in V$ of each edge $e\in E$ as well as all edges
$\del_j(f)\in E$ in the boundary of each face $f\in F$. Here $1\leq
j\leq N(f)$ enumerates all these edges.

As far as the orientations are concerned, we write
$\epsilon(v)\in\{-1,+1\}$ depending on whether the four-simplex dual
to $v\in V$ is isomorphic to a simplex in $M$ or whether this is true
for its counterpart with opposite orientation, denoted by $v^\ast$. For
each $v\in V$, $e\in E$, we write $\epsilon(v,e)\in\{-1,+1\}$ for the
orientation of the tetrahedron dual to $e$ in the boundary of the
four-simplex dual to $v$.

In the interior of $M$, each tetrahedron is contained in the boundary
of exactly two four-simplices so that it appears once with either
orientation. Therefore
\begin{equation}
\label{eq_consist}
  \epsilon(v_1,e)\epsilon(v_1) = -\epsilon(v_2,e)\epsilon(v_2),
\end{equation}
where the tetrahedron dual to $e\in E$ is contained in the boundary of
the two four-simplices dual to $v_1,v_2\in V$.

Finally, we write $\epsilon(e,f)\in\{-1,+1\}$ for the orientation of
the triangle (dual to the face) $f\in F$ in the boundary of the
tetrahedron (dual to the edge) $e\in E$.

In our formulas, we always use the notation of the two-complex
$(V,E,F)$. If we mean the two-complex dual to a triangulation, we
often omit the words `dual to' and speak of the four-simplex $v\in V$,
the tetrahedron $e\in E$, \etc. As the motivation of the causal
Barrett--Crane model involves results from Regge calculus, we are
initially restricted to these two-complexes dual to triangulations. In
the subsequent sections of this article, however, our formulas will be
valid on any oriented two-complex $(V,E,F)$.

%------------------------------------------------------------------------------
\subsection{The original models}
%------------------------------------------------------------------------------

The partition function of the Riemannian or
$\Spin(4)$-symmetric\footnote{Note that we can equally well start from
an $\SO(4)$-symmetry without changing the resulting
model~\cite{Pf02a}. Also note that we consider the version of the
original proposal~\cite{BaCr98} that employs the relativistic
$10j$-symbols.} Barrett--Crane model~\cite{BaCr98} can be
written~\cite{Pf02a},
\begin{eqnarray}
\label{eq_zriemann}
  Z_R &=& \Bigl(\prod_{f\in F}\sum_{j_f\in\frac{1}{2}\N_0}(2j_f+1)\Bigr)\,
    \Bigl(\prod_{e\in E}\int_{S^3}\,dx_e^{(+)}\int_{S^3}\,dx_e^{(-)}\Bigr)\,
    \Bigl(\prod_{f\in F}\sym{A}_f\Bigr)\,
    \Bigl(\prod_{e\in E}\sym{A}_e\Bigr)\nn\\
  &\times& \prod_{v\in V}\Bigl(\prod_{f\in v_0}
    K_R^{(j_f)}(x_{e_+(f,v)}^{(+)},x_{e_-(f,v)}^{(-)})\Bigr),
\end{eqnarray}
where the set $v_0\subseteq F$ includes all faces that contain the
vertex $v\in V$ in their boundary, and $e_+(f,v)\in E$ denotes the
edge in the boundary of the face $f\in F$ that has the vertex
$v=\del_+(e)\in V$ as its target, similarly $e_-(f,v)$. The function,
\begin{equation}
\label{eq_ampriemann}
  K_R^{(j)}(x,y):=\frac{\sin\bigl((2j+1)d_R(x,y)\bigr)}{(2j+1)\sin d_R(x,y)},
\end{equation}
denotes the (normalized) character of $\SU(2)$ in the
$(2j+1)$-dimensional irreducible representation. It depends only on
the relative polar angle
\begin{equation}
\label{eq_angler}
  d_R(x,y):=\cos^{-1}(x\cdot y)\geq 0
\end{equation}
 on $S^3\cong\SU(2)$ where we write normalized vectors $x,y\in
S^3\subseteq\R^4$ and denote by `$\cdot$' the standard scalar product
in $\R^4$.

The model~\eqref{eq_zriemann} contains a sum over all
simple\footnote{We use $\alignidx{V_j\otimes V_j^\ast}$ instead of the
isomorphic $V_j\otimes V_j$ in order to remove all unnecessary signs
from the expressions, see also~\cite{Pf02a,Pf02c}.} (also called balanced)
irreducible representations $\alignidx{V_j\otimes V_j^\ast}$ of
$\SO(4)$ where $V_j\cong\C^{2j+1}$ denotes the irreducible
representations of $\SU(2)$. There are also continuous variables in
the model, namely two integrals over $S^3$ for each edge which
originate from the integral presentation~\cite{Ba98} of the Riemannian
$10j$-symbols. The last product over the $K_R^{(j)}$
in~\eqref{eq_zriemann} is the integrand of the $10j$-symbol. For the
geometric interpretation of the continuous variables, see~\cite{Pf02a}.

The functions $\sym{A}_f$ and $\sym{A}_e$ in the integrand
of~\eqref{eq_zriemann} denote amplitudes for each face and for each
edge which are not fixed by the geometric conditions imposed in the
construction of the Barrett--Crane model~\cite{BaCr98}. There exist
several proposals for these amplitudes in the literature so that we
leave them unspecified in the following calculations.

The Lorentzian Barrett--Crane model whose symmetry group\footnote{By
an analogous argument as before, we could have rather chosen
$\SO_0(1,3)$, the connected component of the Lorentz group that
contains the unit (the proper orthochronous group).} is $\SL(2,\C)$ in
the version in which all tetrahedra are space-like~\cite{BaCr00}, is
defined in complete analogy by the partition function,
\begin{eqnarray}
\label{eq_zlorentz}
  Z_L&=&\Bigl(\prod_{f\in F}\int_0^\infty\,p_f^2\,dp_f\Bigr)\,
    \Bigl(\prod_{e\in E}\int_{H^3_+}\,dx_e^{(+)}\int_{H^3_+}\,dx_e^{(-)}\Bigr)\,
    \Bigl(\prod_{f\in F}\sym{A}_f\Bigr)\,
    \Bigl(\prod_{e\in E}\sym{A}_e\Bigr)\nn\\
  &\times& \prod_{v\in V}\Bigl(\prod_{f\in v_0}
    K_L^{(p_f)}(x_{e_+(f,v)}^{(+)},x_{e_-(f,v)}^{(-)})\Bigr),
\end{eqnarray}
where
\begin{equation}
\label{eq_amplorentz}
  K_L^{(p)}(x,y):=\frac{\sin(p\,d_L(x,y))}{p\sinh d_L(x,y)},
\end{equation}
and 
\begin{equation}
\label{eq_anglel}
  d_L(x,y):=\cosh^{-1}(x\cdot y)\geq 0
\end{equation}
denotes the relative rapidity (hyperbolic distance) of $x,y\in
H^3_+\subseteq\R^{1+3}$. Here we denote by
\begin{equation}
  H^3_+:=\{\,x\in\R^{1+3}\colon\quad x\cdot x=1\quad\mbox{and}\quad x^0>0\,\}
\end{equation}
three-dimensional hyperbolic space, written as the set of future
pointing time-like unit vectors in Mink\-owski space $\R^{1+3}$ whose
standard scalar product $x\cdot y$ is diagonal with entries
$(1,-1,-1,-1)$. Note that we are following the conventions
of~\cite{BaCr00,BaBa01} here. It should be mentioned that the
integrals over $H^3_+$ in~\eqref{eq_zlorentz} converge only after
division by an infinite volume factor~\cite{BaBa01}. We keep this fact
in mind, but leave our formulas unchanged in order to preserve their
full symmetry.

% ==============================================================================
%
\section{The causal models}
%
% ==============================================================================
\label{sect_measure}

%------------------------------------------------------------------------------
\subsection{A construction}
%------------------------------------------------------------------------------
\label{sect_construct}

Livine and Oriti~\cite{LiOr02} have proposed a modification of the
Lorentzian model~\eqref{eq_zlorentz} in which one writes the sine in
the numerator of each factor $K_L^{(p)}$ as a sum of two exponentials
and keeps only one of the exponentials, discarding the other. What
matters is the sign that appears in the exponent. The authors derive
consistency conditions under which one can make this choice for the
entire triangulation so that one obtains a total amplitude $e^{iS}$ in
the integrand in which $S$ is related to the four-dimensional
Lorentzian Regge action. In this section, we present a full
construction of the relevant signs.

We use the notation introduced in Section~\ref{sect_triang} and
observe that the condition~(85) of~\cite{LiOr02} coincides with
our equation~\eqref{eq_consist}.

In order to interpret the model~\eqref{eq_zlorentz} geometrically as
in~\cite{Pf02a}, we wish to associate two future pointing time-like
vectors $x_e^{(\pm)}\in H^3_+$ to each tetrahedron. The only question
is what future pointing means for a purely combinatorial
triangulation.

There is some additional information required, namely a partial order
`$\succ$' on the set $V$ of four-simplices such that the relation is
at least defined for each pair of four-simplices that share a
tetrahedron in their boundary. Without loss of generality, we can then
put only tetrahedra $e$ into the set $E$ for which
$\del_+(e)\succ\del_-(e)$. Otherwise we rather include $e^\ast$ in the
set $E$ and adapt the $\epsilon(\cdot,\cdot)$-factors of
Section~\ref{sect_triang} accordingly.

Observe that if we had triangulated an oriented Lorentzian
four-manifold with only space-like tetrahedra, we could have derived
the partial order `$\succ$' from the causal structure of the metric,
\ie\ `$\succ$' means `is in the causal future of', and the above
construction would precisely result in future pointing time-like
normal vectors.

The choice of signs in~\cite{LiOr02} is based on the Lorentzian Regge
action in four dimensions. Therefore one has to calculate the
`dihedral rapidities' (generalized defect angles) for all pairs of
neighbouring tetrahedra $e_1,e_2\in E$ that cobound a given triangle
$f\in F$. Neighbouring here means that both tetrahedra are contained
in the boundary of the same four-simplex $v\in V$. The outward normals
are then given by
$n_{e_j}:=\epsilon(v,e_j)\epsilon(v)x_{e_j}\in\R^{1+3}$,
$j\in\{1,2\}$, and we also define normals $m_{e_j}=\epsilon(e_j,f)
n_{e_j}$ which go clockwise `around the triangle $f$'. These are used
in order to calculate the Lorentzian analogue of the defect
angle. Recall that all triangles are space-like.

Depending on the shape of the four-simplex $v$, we are either in the
\emph{thin wedge} situation (see~\cite{BaFo94} for the analogous
three-dimensional case), in which the rapidity relevant for the action
corresponds to an interior angle,
\begin{equation}
  \xi(x_{e_1},x_{e_2})=\cosh^{-1}(m_{e_1}\cdot m_{e_2}),
\end{equation}
or else in the \emph{thick wedge} situation in which it corresponds to
an exterior one,
\begin{equation}
  \xi(x_{e_1},x_{e_2})=-\cosh^{-1}(-m_{e_1}\cdot m_{e_2}).
\end{equation}
As all $x_e$ are future pointing, we are in the thin wedge case if and
only if
\begin{equation}
  \epsilon(v,e_1)\epsilon(e_1,f)\epsilon(v,e_2)\epsilon(e_2,f)=+1.
\end{equation}
It can thus be shown that the relevant rapidity is always
\begin{equation}
\label{eq_rapidity}
  \xi(x_{e_1},x_{e_2}):=
    \epsilon(v,e_1)\epsilon(e_1,f)\epsilon(v,e_2)\epsilon(e_2,f)\cosh^{-1}(x_{e_1}\cdot x_{e_2}).
\end{equation}
The contribution to the Lorentzian Regge action from a given triangle
$f\in F$ is therefore,
\begin{equation}
\label{eq_regge}
  S_f = A_f\sum_{v\in f_0}\xi(x^{(+)}_{e_+(f,v)},x^{(-)}_{e_-(f,v)}),
\end{equation}
where $A_f$ denotes the area of the triangle $f\in F$ and the set
$f_0\subseteq V$ contains all four-simplices in whose boundary the
triangle $f$ occurs, so that the sum is over all pairs of tetrahedra
$e_+(f,v)$, $e_-(f,v)$ that share the triangle $f$ and that are
contained in the boundary of the same four-simplex $v\in V$.

While the Regge action $S_f$ for any single triangle is of the
form~\eqref{eq_regge}, each triangle can contribute with a different
total sign $\epsilon(f)\in\{-1,+1\}$ so that the overall action is
rather,
\begin{equation}
\label{eq_regge2}
  S=\kappa\sum_{f\in F}\epsilon(f)S_f,
\end{equation}
where we have also introduced a dimensionless `coupling' constant
$\kappa$.

The $\epsilon(f)$ are not independent though. There exist relations
for each four-simplex from Stokes' theorem for the oriented normal
vectors to the tetrahedra in its boundary which can be evaluated
differentially using the Lorentzian version of the Schl\"afli
identities~\cite{BaSt02}. For each four-simplex $v\in V$ and two
tetrahedra $e_1,e_2\in E$ sharing a common triangle $f\in F$, the
condition is
\begin{equation}
\label{eq_constraint}
  \epsilon(f)=\epsilon(v,e_1)\epsilon(e_1,f)\epsilon(v,e_2)\epsilon(e_2,f)\mu(v),
\end{equation}
where $\mu(v)\in\{-1,+1\}$ denotes an unspecified sign for each
four-simplex. This condition was given in~\cite{BaSt02} writing
$a_e:=\epsilon(v)\epsilon(v,e)\epsilon(e,f)$.

It would now even be possible to absorb the $\mu(v)$ into the
orientation of the four-simplices of the selected triangulation by
putting either $v$ or $v^\ast$ into the set $V$ so that all
$\mu(v)=+1$. It is, however, also possible to keep the orientations of
the four-simplices as they are and to choose $\mu(v):=\epsilon(v)$. As
we will see below, a four-simplex with $\epsilon(v)=+1$ is then
interpreted as a future-pointing contribution to the causal path
integral.

According to~\cite{LiOr02}, the causal Lorentzian Barrett--Crane model
is defined by replacing $K_L^{(p)}(x,y)$ in~\eqref{eq_zlorentz} by,
\begin{equation}
\label{eq_amplcausal}
  {\tilde K}_L^{(p)}(x,y) :=
    \frac{\epsilon(f)e^{i\epsilon(f)p\chi}}{2ip\sinh \chi},\qquad
    \chi:=\epsilon(f)\epsilon(v)d_L(x,y),
\end{equation}
depending on the triangle $f\in F$ and the four-simplex $v\in V$. This
expression leads precisely to the Lorentzian Regge
action~\eqref{eq_regge2} in the exponent. In order to see this,
combine~\eqref{eq_rapidity} with~\eqref{eq_constraint}. We remind the
reader that we are following the conventions of~\cite{BaCr00,BaBa01}
which differ from those employed in~\cite{LiOr02}. By analogy with the
Lorentzian model, we also define a causal Riemannian Barrett--Crane
model replacing $K_R^{(j)}(x,y)$ in~\eqref{eq_zriemann} by,
\begin{equation}
\label{eq_amprcausal}
  {\tilde K}_R^{(j)}(x,y):=
    \frac{\epsilon(f)e^{i\epsilon(f)(2j+1)\phi}}{2i(2j+1)\sin \phi},\qquad
    \phi:=\epsilon(f)\epsilon(v)d_R(x,y).
\end{equation}

%------------------------------------------------------------------------------
\subsection{The measure}
%------------------------------------------------------------------------------

Recall that the integrals over $H^3_+$ in the original version of the
Lorentzian model~\eqref{eq_zlorentz} had to be
regularized~\cite{BaBa01}, exploiting the $\SL(2,\C)$-invariance of
the $10j$-symbol and dividing by an infinite volume factor, in order
to obtain absolutely convergent integrals. It can be seen
from~\eqref{eq_amplcausal} that the same procedure will not suffice in
order to define the corresponding integrals of the causal model
because the overall numerator is of modulus one while the denominator
goes to zero linearly as $\chi\to 0$ so that these integrals will
diverge. Therefore the four-simplex amplitudes cannot even be defined
for a given assignment of representations $p_f$ to the faces.

This situation is different in nature from the expected divergence of
the partition function which is already familiar from the
Ponzano--Regge model in three dimensions, which originates just from
the summation over infinitely many representations and which can be
understood as an infrared divergence~\cite{FrLo02b}.

In order to proceed, we therefore have to modify~\eqref{eq_amplcausal}
and~\eqref{eq_amprcausal} in a suitable way. Let us consider the
Riemannian case first. The first observation is that there is some
freedom in the splitting performed in~\cite{LiOr02} into one factor
which is associated with the overall measure, here the denominator
$(2j+1)\sin\phi$, and another factor, here the numerator
$\sin\bigl((2j+1)\phi\bigr)$, which is interpreted as the amplitude
$e^{iS}\pm e^{-iS}$. The splitting was only motivated by the analogy
with the lower dimensional cases, as outlined, for example,
in~\cite{Ar02}. The key condition is that the amplitude factor can be
written as the sum of two complex conjugate terms of modulus one.

There exists, for example, the following alternative splitting which
leads to a bounded measure part and still satisfies the required
condition on the amplitude part,
\begin{equation}
\label{eq_split}
  \frac{\sin\bigl((2j+1)\phi\bigr)}{(2j+1)\sin\phi}
  = \underbrace{\frac{\sin(\frac{2j+1}{2}\phi)}{(2j+1)\sin\phi}}_{\rm measure}
      \cdot\underbrace{2\cos({\scriptstyle \frac{2j+1}{2}}\phi)}_{\rm amplitude}.
\end{equation}
The cosine of the amplitude part can then still be written as
$e^{iS}+e^{-iS}$. We observe that the numerator of the measure part
goes to zero linearly as $\phi\to 0$ canceling the divergence and at
the same time we get the expression $(j+\frac{1}{2})\phi$ in the
exponent of the amplitude part which uses the area eigenvalue
$(j+\frac{1}{2})$ of~\cite{AlPo00}. At this point the Riemannian model
is very helpful because this area eigenvalue directly motivates the
choice of~\eqref{eq_split}.

The new definitions which replace the causal
amplitudes~\eqref{eq_amplcausal} and~\eqref{eq_amprcausal} are
therefore,
\begin{equation}
\label{eq_amprcausalr}
  K_{R,{\rm causal}}^{(j)}(x,y) :=
    \frac{\sin(\frac{2j+1}{2}\phi)}{(2j+1)\sin\phi}\,e^{i\epsilon(f)(j+\frac{1}{2})\phi},
\end{equation}
and by analogy the Lorentzian case,
\begin{equation}
\label{eq_amplcausalr}
  K_{L,{\rm causal}}^{(p)}(x,y) :=
    \frac{\sin(\frac{p}{2}\chi)}{p\sinh\chi}\,e^{i\epsilon(f)\frac{p}{2}\chi},
\end{equation}
where $\phi$ and $\chi$ are defined as in~\eqref{eq_amplcausal}
and~\eqref{eq_amprcausal}.

In the Riemannian case, the integrals over the $x_e^{(\pm)}\in S^3$
converge because all $K^{(j)}_{R,{\rm causal}}$ are bounded and
integrated over a compact manifold $S^3\times\cdots\times S^3$. In the
Lorentzian case, we notice that
\begin{equation}
  |K_{L,{\rm causal}}^{(p)}(x,y)|\leq2|K_L^{(\frac{p}{2})}(x,y)|,
\end{equation}
so that the proof of convergence of the integrals of $K_L^{(p)}(x,y)$
over the $x_e^{(\pm)}\in H^3_+$ which was presented in~\cite{BaBa01},
implies the existence of our integrals of the $K_{L,{\rm
causal}}^{(p)}$ over the $x_e^{(\pm)}\in H^3_+$. We can use the same
regularization procedure.

%------------------------------------------------------------------------------
\subsection{A coupling constant}
%------------------------------------------------------------------------------
\label{sect_constant}

In our notation, we can simplify~\eqref{eq_amprcausalr}
and~\eqref{eq_amplcausalr} as follows,
\begin{eqnarray}
\label{eq_amplcausal2}
  K_{L,{\rm causal}}^{(p)}(x,y) &=&
    \epsilon(v)\epsilon(f)\frac{\sin\bigl(\frac{p}{2}d_L(x,y)\bigr)}{p\sinh d_L(x,y)}\,
      e^{i\epsilon(v)\,\frac{p}{2}d_L(x,y)},\\
\label{eq_amprcausal2}
  K_{R,{\rm causal}}^{(j)}(x,y) &=&
    \epsilon(v)\epsilon(f)\frac{\sin\bigl(\frac{2j+1}{2}d_R(x,y)\bigr)}{(2j+1)\sin d_R(x,y)}\,
      e^{i\epsilon(v)\,(j+\frac{1}{2})d_R(x,y)},
\end{eqnarray}
so that the exponents depend only on one sign $\epsilon(v)$ per
four-simplex. This links the orientation of four-simplices with the
amplitudes of the causal model. Therefore we can write the partition
function of the causal Lorentzian Barrett--Crane model as,
\begin{eqnarray}
\label{eq_zclorentz}
  Z_{L,{\rm causal}}
  &=&\Bigl(\prod_{f\in F}\int_0^\infty\,p_f^2\,dp_f\Bigr)\,
     \Bigl(\prod_{e\in E}\int_{H^3_+}\,dx_e^{(+)}\int_{H^3_+}\,dx_e^{(-)}\Bigr)\,
     \Bigl(\prod_{f\in F}\sym{A}_f\Bigr)\,
     \Bigl(\prod_{e\in E}\sym{A}_e\Bigr)\nn\\
  &\times& \prod_{v\in V}\Bigl(\prod_{f\in v_0}
     \frac{\sin\bigl(\frac{p_f}{2}d_L(x_{e_+(f,v)}^{(+)},x_{e_-(f,v)}^{(-)})\bigr)}
          {p_f\sinh d_L(x_{e_+(f,v)}^{(+)},x_{e_-(f,v)}^{(-)})}\Bigr)\,
     \exp(iS_L),
\end{eqnarray}
where
\begin{equation}
\label{eq_reggel}
  S_L := \kappa\sum_{v\in V}\epsilon(v)\sum_{f\in v_0}\frac{p_f}{2}d_L(x_{e_+(f,v)}^{(+)},x_{e_-(f,v)}^{(-)})
\end{equation}
denotes the Regge action in the variables of the Lorentzian model in
which the areas of the triangles $f\in F$ are given by the $p_f/2\geq
0$. Observe that the prefactors $\epsilon(v)\epsilon(f)$
of~\eqref{eq_amplcausal2} cancel in the product over all vertices and
all faces attached to the vertices.

The causal Riemannian model is given by,
\begin{eqnarray}
\label{eq_zcriemann}
  Z_{R,{\rm causal}}
  &=& \Bigl(\prod_{f\in F}\sum_{j_f\in\frac{1}{2}\N_0}(2j_f+1)\Bigr)\,
      \Bigl(\prod_{e\in E}\int_{S^3}\,dx_e^{(+)}\int_{S^3}\,dx_e^{(-)}\Bigr)\,
      \Bigl(\prod_{f\in F}\sym{A}_f\Bigr)\,
      \Bigl(\prod_{e\in E}\sym{A}_e\Bigr)\nn\\
  &\times& \prod_{v\in V}\Bigl(\prod_{f\in v_0}
     \frac{\sin\bigr(\frac{2j_f+1}{2}d_R(x_{e_+(f,v)}^{(+)},x_{e_-(f,v)}^{(-)})\bigl)}
          {(2j_f+1)\sin d_R(x_{e_+(f,v)}^{(+)},x_{e_-(f,v)}^{(-)})}\Bigr)\,
     \exp(iS_R),
\end{eqnarray}
where
\begin{equation}
\label{eq_regger}
  S_R := \kappa\sum_{v\in V}\epsilon(v)\sum_{f\in v_0}\bigl(j_f+\frac{1}{2}\bigr)d_R(x_{e_+(f,v)}^{(+)},x_{e_-(f,v)}^{(-)})
\end{equation}
is the Regge action in terms of the variables of the Riemannian model
in which the triangle areas are given by $j_f+\frac{1}{2}$.

In both cases, the amplitudes of the causal model lead to the Regge
action for the special value $\kappa=1$ which in turn indicates that
there could have been be a free parameter $\kappa$ in the
Barrett--Crane model right from the beginning. This observation
suggests the following generalization of the original
amplitudes~\eqref{eq_ampriemann} and~\eqref{eq_amplorentz} to
\begin{eqnarray}
  K_R^{(j)}(x,y)&:=&
    \frac{\sin\bigl(\frac{2j+1}{2}d_R(x,y)\bigr)}{(2j+1)\sin d_R(x,y)}
    \cdot 2\cos\bigl(\frac{2j+1}{2}\kappa\,d_R(x,y)\bigr),\\
\label{eq_amploriginal}
  K_L^{(p)}(x,y)&:=&
    \frac{\sin\bigl(\frac{p}{2}d_L(x,y)\bigr)}{p\sinh d_L(x,y)}
    \cdot 2\cos\bigl(\frac{p}{2}\kappa\,d_L(x,y)\bigr),
\end{eqnarray}
and of the causal amplitudes~\eqref{eq_amplcausal2}
and~\eqref{eq_amprcausal2} to
\begin{eqnarray}
\label{eq_amplcausal3}
  K_{L,{\rm causal}}^{(p)}(x,y) &=&
    \epsilon(v)\epsilon(f)\frac{\sin\bigl(\frac{p}{2}d_L(x,y)\bigr)}{p\sinh d_L(x,y)}\,
      e^{i\epsilon(v)\,\frac{p}{2}\kappa\,d_L(x,y)},\\
  L_{R,{\rm causal}}^{(j)}(x,y) &=&
    \epsilon(v)\epsilon(f)\frac{\sin\bigl(\frac{2j+1}{2}d_R(x,y)\bigr)}{(2j+1)\sin d_R(x,y)}\,
      e^{i\epsilon(v)\,(j+\frac{1}{2})\kappa\,d_R(x,y)}.
\end{eqnarray}
Observe that we have inserted the coupling constant $\kappa$ only into
the amplitude part, but not into the measure part, \cf~\eqref{eq_split}.

Once one has accepted the idea for the construction of the causal
model, some splitting such as~\eqref{eq_split} is necessary in order
to make the integrals over $H^3_+$ or $S^3$ convergent. The constant
$\kappa$ parametrizes in some sense the non-uniqueness of such a
splitting. Notice that the original model for $\kappa\neq 1$ no longer
satisfies the bi-vector constraints of~\cite{BaCr98} nor does the
causal model satisfy them.

At this point, the various models diverge from each other, and the
important question is what the physical relevance of the
splitting~\eqref{eq_split} and of the constant $\kappa$ is. In the
following sections, we provide some tools in order to study these
questions. These are first the definition of an Euclidean version of
the causal model in order to apply numerical simulations and second
the study of the symmetries and the most general observables of these
models.

As far as the significance of the constant $\kappa$ is concerned,
there seem to prevail two opposite philosophies among the experts. If
a classical limit can be obtained by studying the large spin (or
semi-classical) limit for a single four-simplex, the appearance of
$\kappa$ is unlikely to have much impact. If, however, the
four-simplices are too strongly coupled and the classical limit
requires a non-perturbative renormalization by `block-spin' or coarse
graining transformations in the spirit of Statistical Mechanics, then
the new parameter $\kappa$ can easily influence the results, even in
the original Barrett--Crane model. For numerical results in the
original model, see~\cite{BaCh02c}. Even the fact that the original
Barrett--Crane model with $\kappa\neq1$ does not satisfy the bi-vector
constraints anymore, would not necessarily be fatal. If $\kappa$
happens to control the renormalization scale, it is conceivable that
the constraints are satisfied only approximately at an effective
coarse grained scale.

It was already observed in~\cite{LiOr02} that the variables of the
path integral, independent areas $p_f$ or $j_f$ and directions
$x^{(\pm)}_e\in H^3_+$ or $S^3$, indicate that the Regge
actions~\eqref{eq_reggel} and~\eqref{eq_regger} have to be understood
as actions in a first order formalism~\cite{Ba94}. This is consistent
with the fact that the Regge action appears only in the intermediate
stage of the duality transformation~\cite{Pf02a} in which both types
of variables are present. The necessary constraint~\cite{Ba94} that
the variation of the angles is only over dihedral angles that
correspond to actual four-simplex geometries, has been automatically
implemented in the construction of the causal model, see
Section~\ref{sect_construct} or~\cite{LiOr02}.

% ==============================================================================
%
\section{Euclideanization}
%
% ==============================================================================
\label{sect_wick}

In both partition functions~\eqref{eq_zclorentz}
and~\eqref{eq_zcriemann}, the Regge action takes its simplest form,
\ie\ with the least number of explicit signs, if one sums over all
four-simplices only in the last step. If one interprets each
four-simplex with $\epsilon(v)=+1$ as a future-pointing contribution
to the causal path integral, \ie\ obtained after integrating only over
positive lapse, for details see~\cite{LiOr02}, there is an obvious
candidate for an Euclidean model by a suitable modification of the
amplitudes. This can be done four-simplex by four-simplex. If we
substitute $\epsilon(v)\cdot i$ into all exponents, we turn the
oscillations into an exponential damping and arrive at the following
proposal for an Euclidean Lorentzian Barrett--Crane model,
\begin{eqnarray}
\label{eq_zelorentz}
  Z_{L,{\rm Eucl.}}
  &=&\Bigl(\prod_{f\in F}\int_0^\infty\,p_f^2\,dp_f\Bigr)\,
     \Bigl(\prod_{e\in E}\int_{H^3_+}\,dx_e^{(+)}\int_{H^3_+}\,dx_e^{(-)}\Bigr)\,
     \Bigl(\prod_{f\in F}\sym{A}_f\Bigr)\,
     \Bigl(\prod_{e\in E}\sym{A}_e\Bigr)\nn\\
  &\times& \prod_{v\in V}\Bigl(\prod_{f\in v_0}
     \frac{\sin\bigl(\frac{p_f}{2}d_L(x_{e_+(f,v)}^{(+)},x_{e_-(f,v)}^{(-)})\bigr)}
          {p_f\sinh d_L(x_{e_+(f,v)}^{(+)},x_{e_-(f,v)}^{(-)})}\Bigr)\,
     \exp(-S_{L,{\rm Eucl.}}),
\end{eqnarray}
where
\begin{equation}
  S_{L,{\rm Eucl.}} := \kappa\sum_{v\in V}\sum_{f\in v_0}\frac{p_f}{2}d_L(x_{e_+(f,v)}^+,x_{e_-(f,v)}^-).
\end{equation}
This means that we have replaced $K_{L,{\rm causal}}^{(p)}$ by
\begin{equation}
\label{eq_ampleuclidean}
  K_{L,{\rm Eucl.}}^{(p)}(x,y):=\frac{\sin\bigl(\frac{p}{2}d_L(x,y)\bigr)}{p\sinh d_L(x,y)}
    e^{-\frac{p}{2}\kappa d_L(x,y)}.
\end{equation}
Similarly for Riemannian signature,
\begin{eqnarray}
\label{eq_zeriemann}
  Z_{R,{\rm Eucl.}}
  &=& \Bigl(\prod_{f\in F}\sum_{j_f\in\frac{1}{2}\N_0}(2j_f+1)\Bigr)\,
      \Bigl(\prod_{e\in E}\int_{S^3}\,dx_e^{(+)}\int_{S^3}\,dx_e^{(-)}\Bigr)\,
      \Bigl(\prod_{f\in F}\sym{A}_f\Bigr)\,
      \Bigl(\prod_{e\in E}\sym{A}_e\Bigr)\nn\\
  &\times& \prod_{v\in V}\Bigl(\prod_{f\in v_0}
     \frac{\sin\bigl(\frac{2j_f+1}{2}d_R(x_{e_+(f,v)}^{(+)},x_{e_-(f,v)}^{(-)})\bigr)}
          {(2j_f+1)\sin d_R(x_{e_+(f,v)}^{(+)},x_{e_-(f,v)}^{(-)})}\Bigr)\,
     \exp(-S_{R,{\rm Eucl.}}),
\end{eqnarray}
where
\begin{equation}
  S_{R,{\rm Eucl.}} := \kappa\sum_{v\in V}\sum_{f\in v_0}\bigl(j_f+\frac{1}{2}\bigr)d_R(x_{e_+(f,v)}^+,x_{e_-(f,v)}^-),
\end{equation}
\ie\ we have replaced $K_{R,{\rm causal}}^{(j)}$ by
\begin{equation}
  K_{R,{\rm Eucl.}}^{(j)}(x,y) := \frac{\sin\bigl(\frac{2j+1}{2}d_R(x,y)\bigr)}{(2j+1)\sin d_R(x,y)}
    e^{-(j+\frac{1}{2})\kappa d_R(x,y)}.
\end{equation}
We note that $S_{L,{\rm Eucl.}}\geq 0$ and $S_{R,{\rm Eucl.}}\geq 0$
for any $\kappa\geq 0$.

Unfortunately, the Euclidean reconstruction~\cite{AsMa00} does not
provide a recipe of how to derive the \emph{Euclidean action}, \ie\
how to choose the four-dimensional path integral measure. In a fixed
background with a distinguished $t$-coordinate, one is usually guided
by the heuristic substitution $t\mapsto -i\tau$ which typically
changes some signs in the action wherever there are time derivatives
involved, \ie\ in the kinetic energy part. With the special `time'
coordinate used in three-dimensional Lorentzian dynamical
triangulations, there is still a similar substitution one can use in
order to construct the Euclidean action~\cite{AmJu01}. The
substitution used in those models essentially distinguishes space-like
from time-like simplices and introduces a relative sign.

In the case of spin foam models, we do not have such a `time'
coordinate at hand. However, since in the Lorentzian case all
triangles are space-like and therefore treated on equal footing, we
expect that no relative signs enter the Euclidean action. As long as
we capture the relevant local symmetries, one can expect that Euclidean
reconstruction will lead to the correct canonical theory by
universality arguments.

We observe that if we fix all representations associated to the faces
to the same representation, we get regular flat simplices so that we
are in a situation very similar to a single configuration in a
dynamical triangulation model~\cite{AmJu01}. The main difference to
the dynamical triangulation models is that in those models all
simplices have the same geometry, in particular the same size, and
that all dynamical properties of the geometry are encoded in the sum
over triangulations. In the case of the Barrett--Crane models, we have
the additional complication that the geometry of the individual
simplices is determined by the assignment of representations $j_f$ or
$p_f$ to the triangles. Therefore, the individual simplices can
already be arbitrarily large. We also stress that the
formulas~\eqref{eq_zelorentz} and~\eqref{eq_zeriemann} refer to a
two-complex with a causal structure imposed on the vertices. This
excludes in particular space-times with closed time-like curves. Even
stronger, if we have Euclidean reconstruction following~\cite{AsMa00 }
in mind, we cannot yet deal with topology change and have to require a
global space-time topology of $\Sigma\times\R$.

In the following section, we study expectation values of the
continuous variables of the Barrett--Crane models. With these
definitions, one can easily establish a dictionary in order to compare
our constructions with the generalized Euclidean reconstruction
of~\cite{AsMa00}. Note that the Euclidean measure of~\cite{AsMa00}
includes the integration measures, the factors which we have called
the \emph{measure part} and also what we have called the
\emph{amplitude part}. On very regular triangulations, it is possible
to check under which conditions the Euclidean path integral measure of
the Barrett--Crane model is reflection positive (this refers to what
is called \emph{link reflection positive} by Lattice Gauge Theorists).
This condition is satisfied provided that the edge and face amplitudes
are real, \ie\ if they do not change when one dualizes all
representations involved.

% ==============================================================================
%
\section{Observables}
%
% ==============================================================================
\label{sect_observe}

In this section, we consider all Barrett--Crane models mentioned so
far as path integrals over the continuous connection variables
$x^{(\pm)}_e\in S^3$ or $H^3_+$. This point of view has already been
adopted in~\cite{Pf02a} for the Riemannian model where we have shown
that one can perform the sums over the representations
$j=0,\frac{1}{2},1,\ldots$ as soon as the edge amplitudes are
sufficiently simple. This point of view is also more closely related
to the Euclidean reconstruction~\cite{AsMa00} than is the usual
picture in which the variables of the path integral are the
representations attached to the triangles. In the following, we
therefore consider the $x^{(\pm)}_e$ as the variables of the path
integral while the rest of the partition function, including the sums
over the $j_f$ or the integrals over the $p_f$, belongs to the
amplitudes.

%------------------------------------------------------------------------------
\subsection{Local symmetries}
%------------------------------------------------------------------------------
\label{sect_local}

All the versions of the Riemannian signature model, the original
one~\eqref{eq_zriemann}, the causal~\eqref{eq_zcriemann} and the
Euclidean one~\eqref{eq_zeriemann}, are invariant under the following
local $\Spin(4)$ (or $\SO(4)$) symmetry~\cite{Pf02a},
\begin{eqnarray}
\label{eq_localr}
  x_e^{(+)} &\mapsto& h_{\del_+(e)}x^{(+)}_e{\tilde h}_{\del_+(e)}^{-1},\nn\\
  x_e^{(-)} &\mapsto& h_{\del_-(e)}x^{(-)}_e{\tilde h}_{\del_-(e)}^{-1},
\end{eqnarray}
where $(h_v,\tilde h_v)\in\Spin(4)$, for all $v\in V$, defines a
generating function of this local gauge transformation. We have
identified $S^3\cong\SU(2)$, and the products in~\eqref{eq_localr} are
in $\SU(2)$. The independence follows from the invariance of the
scalar product $x\cdot y$ in $\R^4$ in the definition of $d_R(x,y)$,
\cf~\eqref{eq_angler}.

The Lorentzian counterpart of this local symmetry is given by,
\begin{eqnarray}
\label{eq_locall}
  x_e^{(+)} &\mapsto& h_{\del_+(e)}\cdot x^{(+)}_e,\nn\\
  x_e^{(-)} &\mapsto& h_{\del_-(e)}\cdot x^{(-)}_e,
\end{eqnarray}
where $h_v\in\SL(2,\C)$ for each $v\in V$, and the dot denotes the
action of $\SL(2,\C)$ on Minkowski space $\R^{1+3}$. Again, this
symmetry is a consequence of the invariance of the scalar product in
Minkowski space under the action of $\SL(2,\C)$ which appears in the
definition of $d_L(x,y)$, \cf~\eqref{eq_anglel}. All versions of the
Lorentzian signature model, \eqref{eq_zlorentz}, \eqref{eq_zclorentz}
and~\eqref{eq_zelorentz} are invariant under~\eqref{eq_locall}.

%------------------------------------------------------------------------------
\subsection{Most general expectation values}
%------------------------------------------------------------------------------

The most general functions of the variables $x_e^{(\pm)}$ that are
invariant under these local transformations, can be calculated by
standard techniques (see, for example,~\cite{Pf02b} for detailed
examples). An orthonormal basis for such functions is characterized by
$\Spin(4)$ or $\SL(2;\C)$ spin networks on the graph $(V,E)$.

For the Riemannian case, let $\ell_e=0,\frac{1}{2},1,\ldots$ specify a
simple irreducible representation $\alignidx{V_\ell\otimes
V_\ell^\ast}$ of $\Spin(4)$ for each edge $e\in E$, and let
\begin{equation}
\label{eq_interr}
  P^{(v)}\colon
    \Bigl(\bigotimes_{\ontop{e\in E\colon}{v=\del_-(e)}}{(\alignidx{V_{\ell_e}\otimes V_{\ell_e}^\ast})}^\ast\Bigr)
    \otimes
    \Bigl(\bigotimes_{\ontop{e\in E\colon}{v=\del_+(e)}}(\alignidx{V_{\ell_e}\otimes V_{\ell_e}^\ast})\Bigr)
    \to\C
\end{equation}
denote (an arbitrary, but suitably normalized) $\Spin(4)$-intertwiner
for each vertex $v\in V$. Then the spin network function,
\begin{eqnarray}
\label{eq_spinnetr}
  F_{\ell,P}(\{x_e^{(\pm)}\}) &=&
  \prod_{v\in V}\Biggl[
    \Bigl(\prod_{\ontop{e\in E\colon}{v=\del_+(e)}}\sum_{p_e,q_e=1}^{2\ell_e+1}\Bigr)
    \Bigl(\prod_{\ontop{e\in E\colon}{v=\del_-(e)}}\sum_{r_e,s_e=1}^{2\ell_e+1}\Bigr)\nn\\
    &\times&\Bigl(\prod_{\ontop{e\in E\colon}{v=\del_+(e)}}t^{(\ell_e)}_{p_eq_e}(x_e^{(+)})\Bigr)
    \Bigl(\prod_{\ontop{e\in E\colon}{v=\del_-(e)}}\overline{t^{(\ell_e)}_{r_es_e}(x_e^{(-)})}\Bigr)
    P^{(v)}_{\underbrace{\scriptstyle (r_es_e)\ldots}_{\ontop{e\in E\colon}{v=\del_-(e)}}
             \underbrace{\scriptstyle (p_eq_e)\ldots}_{\ontop{e\in E\colon}{v=\del_+(e)}}}\Biggr],
\end{eqnarray}
is invariant under the local symmetry~\eqref{eq_localr}. The
$t_{pq}^{(\ell)}$ are the representation functions of $\SU(2)\cong
S^3$ and we follow the conventions of~\cite{Pf02a}. Any $L^2$-function
of the $x_e^{(\pm)}$ that is invariant under the local symmetry, is a
square summable series over spin network functions of the
form~\eqref{eq_spinnetr}.

For the Lorentzian case, let $V_{(0,q_e)}$, $q_e\geq 0$, denote a
simple irreducible representation of $\SL(2,\C)$ for each edge $e\in
E$. The vectors of these representation spaces can be modeled by
functions $H^3_+\to\C$, see~\cite{BaCr00,BaBa01} for
details. Employing the Gel'fand--Graev transform, an orthonormal basis
for $V_{(0,q)}$ is given by the functions,
\begin{equation}
\label{eq_gelfand}
  H_{jm}^{(q)}\colon H^3_+\to\C,\qquad x\mapsto
    H_{jm}^{(q)}(x):=\int_\Gamma Y_{jm}(\xi){(x\cdot\xi)}^{-1-ip}\,d\xi,
\end{equation}
where $\Gamma$ denotes the two-sphere of future-pointing light-like
vectors whose spatial component are unit vectors, and the integral is
performed using the normalized Lebesgue measure of $\Gamma$. The
indices of the spherical harmonics $Y_{jm}$ are in the range
$j=0,1,2,\ldots$ and $-j\leq m\leq j$. Let furthermore
\begin{equation}
\label{eq_interl}
  Q^{(v)}\colon 
    \bigl(\bigotimes_{\ontop{e\in E\colon}{v=\del_-(e)}}V_{(0,q_e)}^\ast\bigr)
    \otimes
    \bigl(\bigotimes_{\ontop{e\in E\colon}{v=\del_+(e)}}V_{(0,q_e)}\bigr)
    \to\C
\end{equation}
denote (an arbitrary) $\SL(2,\C)$-intertwiner for each vertex $v\in
V$, given in terms of the coefficients,
\begin{equation}
  Q^{(v)}_{
  \underbrace{\scriptstyle (k_en_e),\ldots,}_{\ontop{e\in E\colon}{v=\del_-(e)}}
  \underbrace{\scriptstyle (j_em_e),\ldots}_{\ontop{e\in E\colon}{v=\del_+(e)}}}
\end{equation}
with respect to the above basis. Then the spin network function
\begin{eqnarray}
\label{eq_spinnetl}
  G_{q,Q}(\{x_e^{(\pm)}\}) &=&
  \prod_{v\in V}\Biggl[
    \Bigl(\prod_{\ontop{e\in E\colon}{v=\del_+(e)}}\sum_{j_e=0}^\infty\sum_{m_e=-j_2}^{j_e}\Bigr)
    \Bigl(\prod_{\ontop{e\in E\colon}{v=\del_-(e)}}\sum_{k_e=0}^\infty\sum_{n_e=-k_e}^{k_e}\Bigr)\\
    &\times&\Bigl(\prod_{\ontop{e\in E\colon}{v=\del_+(e)}}H^{(q_e)}_{j_em_e}(x_e^{(+)})\Bigr)
            \Bigl(\prod_{\ontop{e\in E\colon}{v=\del_-(e)}}\overline{H^{(q_e)}_{k_en_e}(x_e^{(-)})}\Bigr)
    Q^{(v)}_{\underbrace{\scriptstyle (k_en_e)\ldots}_{\ontop{e\in E\colon}{v=\del_-(e)}}
             \underbrace{\scriptstyle (j_em_e)\ldots}_{\ontop{e\in E\colon}{v=\del_+(e)}}}\Biggr],\nn
\end{eqnarray}
is invariant under the local symmetry~\eqref{eq_locall}. All
$L^2$-functions of the $x_e^{(\pm)}$ that are invariant under this
local symmetry, are Plancherel integrals over spin network functions
of the form~\eqref{eq_spinnetl}. There will, however, arise
convergence issues similar to those studied in~\cite{BaBa01}.

If one views the partition function of the Barrett--Crane model as a
path integral over the continuous variables $x_e^{(\pm)}$, the numbers
one can extract from the model are precisely the expectation values of
spin network functions of the form~\eqref{eq_spinnetr}
or~\eqref{eq_spinnetl}, respectively. In the Riemannian case, these
expectation values read
\begin{eqnarray}
\label{eq_expectr}
  \left<F_{\ell,P}\right>
  &=& \frac{1}{Z_{R,X}}\Bigl(\prod_{e\in E}\int_{S^3}\,dx_e^{(+)}\int_{S^3}\,dx_e^{(-)}\Bigr)\,
      F_{\ell,P}(\{x_e^{(\pm)}\})\,\Bigl(\prod_{f\in F}\sum_{j_f\in\frac{1}{2}\N_0}(2j_f+1)\Bigr)\nn\\
  &\times&\Bigl(\prod_{f\in F}\sym{A}_f\Bigr)\,
          \Bigl(\prod_{e\in E}\sym{A}_e\Bigr)\,
          \prod_{v\in V}\Bigl(\prod_{f\in v_0}
    K_{R,X}^{(j_f)}(x_{e_+(f,v)}^{(+)},x_{e_-(f,v)}^{(-)})\Bigr),
\end{eqnarray}
for spin network functions $F_{\ell,P}$ of the
form~\eqref{eq_spinnetr}. Here the symbol $X$ in $Z_{R,X}$ and
$K_{R,X}$ stands for `original', `causal' or `Euclidean',
respectively.

In the Lorentzian case, the analogous expectation value reads
\begin{eqnarray}
\label{eq_expectl}
  \left<G_{q,Q}\right>
  &=&\frac{1}{Z_{L,X}}\Bigl(\prod_{e\in E}\int_{H^3_+}\,dx_e^{(+)}\int_{H^3_+}\,dx_e^{(-)}\Bigr)\,
     G_{q,Q}(\{x_e^{(\pm)}\})\,\Bigl(\prod_{f\in F}\int_0^\infty\,p_f^2\,dp_f\Bigr)\nn\\
  &\times&\Bigl(\prod_{f\in F}\sym{A}_f\Bigr)\,
          \Bigl(\prod_{e\in E}\sym{A}_e\Bigr)\,
          \prod_{v\in V}\Bigl(\prod_{f\in v_0}
    K_{L,X}^{(p_f)}(x_{e_+(f,v)}^{(+)},x_{e_-(f,v)}^{(-)})\Bigr).
\end{eqnarray}
The Euclidean reconstruction~\cite{AsMa00} relies on this type of
expectation values in the construction of the physical Hilbert space.

The expressions~\eqref{eq_expectr} and~\eqref{eq_expectl} can be
reformulated in the language of a path integral whose variables are
representations assigned to the faces and in which the integrals over
the $x_e^{(\pm)}$ are performed, resulting in relativistic
$10j$-symbols as the vertex amplitudes. We call this formulation in
which the representations $j_f$ or $p_f$ are the variables of the path
integral, the \emph{representation picture} as opposed to the
\emph{connection picture} in which the continuous variables
$x_e^{(\pm)}$ are the variables of the path integral. The
transformation from one to the other picture proceeds in complete
analogy to the calculation for the partition function presented
in~\cite{Pf02a}.

We can therefore re-express the expectation values~\eqref{eq_expectr}
and~\eqref{eq_expectl} in the representation picture. In the case of
the original models~\eqref{eq_zriemann} and~\eqref{eq_zlorentz} with
$\kappa=1$, the result takes the simplest form. If, say in the
Riemannian version, the spin network function $F_{\ell,P}$ is
supported only on edges in the boundary of the two-complex, the
expectation value $\left<F_{\ell,P}\right>$ agrees with a matrix
element of spin network states. This means it is calculated by summing
over all spin foams of the Barrett--Crane model living on the given
two-complex, but with additional faces and edges added so that these
faces and edges are coloured by the same representations and
intertwiners $(\{\ell_e\},\{P^{(v)}\})$ as those that characterize the
spin network function $F_{\ell,P}$. These expressions are the desired
matrix elements between spin network states. A completely analogous
result holds for the original Lorentzian model.

Observe that the expectation values~\eqref{eq_expectr}
and~\eqref{eq_expectl} are more general than just such matrix elements
of spin network states. First, we have shown that the intertwiners
$P^{(v)}$ and $Q^{(v)}$ can be generic $\Spin(4)$ or
$\SL(2,\C)$-intertwiners and are not restricted to the special
Barrett--Crane intertwiners. The standard conjecture seems to be that
for pure gravity, it is sufficient to employ \emph{relativistic} spin
networks, \ie\ spin networks whose representations are simple and
whose intertwiners are the Barrett--Crane intertwiner. If this happens
to be true, then our calculation above parametrizes the most generic
way of coupling other fields to pure gravity. This will be of
relevance when one studies the coupling of matter to the
Barrett--Crane model. Indeed the choice of
intertwiners~\eqref{eq_interr} and~\eqref{eq_interl} is the first
occasion where the difference of $\Spin(4)$ or $\SL(2,\C)$ versus
$\SO(4)$ or $\SO_0(1,3)$ matters.

The second aspect in which the expectation values~\eqref{eq_expectr}
and~\eqref{eq_expectl} are more general than matrix elements of spin
network states, is the fact that they are not restricted to the
boundary of the four-manifold. This can be seen as an analogy to the
Wilson loop in Lattice Gauge Theory which is used to determine the
static potential between a quark-antiquark pair. This loop is not only
supported on the space-like boundary of the four manifold, but it
extends in time direction in the interior. This construction serves as
a simplified version of a matter coupling which captures only the
colour properties of the matter field but which neglects its
dynamics. Similar constructions may also prove useful in the study of
spin foam models of quantum gravity. 

In particular, a generic Wilson loop in the connection variables will
give access to the curvature of the full $\Spin(4)$ or
$\SL(2,\C)$-connection and therefore to dynamical properties and not
just to its restriction to a space-like boundary.

Finally, we stress that the study of expectation values such as
$\left<F_{\ell,P}\right>$ and $\left<G_{q,Q}\right>$ is a convenient
way of sidestepping the often tedious technicalities when one deals
with boundary terms.

% ==============================================================================
%
\section{Technical issues}
%
% ==============================================================================
\label{sect_back}

%------------------------------------------------------------------------------
\subsection{Back to the $10j$-symbols}
%------------------------------------------------------------------------------
\label{sect_transback}

The expectation values $\left<F_{\ell,P}\right>$ and
$\left<G_{q,Q}\right>$ become more complicated than what we have
discussed so far, as soon as we consider the causal or the Euclidean
model or $\kappa\neq 1$ in the original model. Recall that in the
original partition function~\eqref{eq_zlorentz}, the integrations over
the $H^3_+$ together with the product of $K_L^{(p)}$ form the
relativistic $10j$-symbols (up to the regularization mentioned
above). As soon as we replace the $K_L^{(p)}$
by~\eqref{eq_amploriginal}, \eqref{eq_amplcausal3}
or~\eqref{eq_ampleuclidean}, we no longer have a model whose
four-simplex amplitudes are the relativistic $10j$-symbols. In the
original model for $\kappa=1$, we were able to `solve' the
integrations over the $H^3_+$ and knew that the result of the
integration had an abstract definition as a relativistic
$10j$-symbol. After the modification of the integrands
$K_{L,X}^{(p)}$, there is no obvious analogy available.

In the following we show how the local symmetry of
Section~\ref{sect_local} can be exploited in order to expand the
modified integrand, a product of factors $K_{L,X}^{(p)}$, into a
series of ordinary relativistic $10j$-symbols. The novel feature is
that this step requires an additional colouring of the \emph{wedges}
(the intersection of a face of the two-complex dual to the
triangulation with a four-simplex of the original triangulation) with
simple representations of the symmetry group.

Consider first the Riemannian case. The functions $K_{R,X}^{(j)}(x,y)$
where $X$ stands for `original', `causal' or `Euclidean', are
$L^2$-functions $S^3\times S^3\to\C$. Since they depend only on $\cos
d_R(x,y)=\frac{1}{2}\chi^{(\frac{1}{2})}(\alignidx{g_x\cdot g_y^{-1}})$, where
$g_x,g_y\in\SU(2)$ denote the corresponding elements of $\SU(2)\cong
S^3$, they are class functions on $\SU(2)$ and can be character
expanded into a square summable series,
\begin{equation}
\label{eq_character}
  K_{R,X}^{(j)}(x,y)=\sum_{k=0,\frac{1}{2},1,\ldots}
    \hat K_k^{(j)}\,\chi^{(k)}(\alignidx{g_x\cdot g_y^{-1}}),\qquad
  \hat K_k^{(j)}:=\int_{\SU(2)}\overline{\chi^{(k)}(g)}\,K_{R,X}^{(j)}(g)\,dg,
\end{equation}
where we write $K^{(j)}_{R,X}(g)$ in order to indicate that
$K^{(j)}_{R,X}(x,y)$ is a function of $g=\alignidx{g_x\cdot g_y^{-1}}$.

\begin{figure}[t]
\begin{center}
\epsfig{file=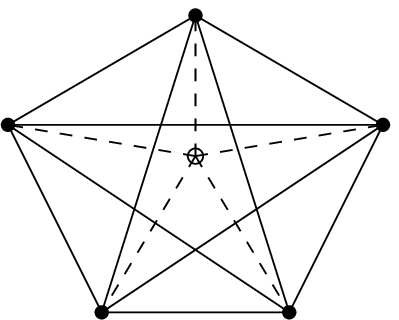}
\end{center}
\mycaption{fig_10j}{% 
The four-simplex amplitude of the expectation value
$\left<F_{\ell,P}\right>$ in~\eqref{eq_backriemann} is the
relativistic $10j$-symbol (solid lines) in which a piece of the spin
network $(\ell,P)$ has been inserted (dashed lines). The full dots
denote the Barrett--Crane intertwiners and the white dot the
intertwiner $P$ of the spin network.}
\end{figure}

With this expansion performed for each triangle $f\in F$ and all
four-simplices $v\in f_0\subseteq F$ that contain the triangle $f$ in
their boundary, we can apply the techniques of~\cite{Pf02a} and
obtain,
\begin{eqnarray}
\label{eq_backriemann}
  \left<F_{\ell,P}\right>&=&\frac{1}{Z_{R,X}}
    \Bigl(\prod_{f\in F}\sum_{j_f=0,\frac{1}{2},1,\ldots}\Bigr)\,
    \Bigl(\prod_{f\in F}\prod_{v\in f_0}\sum_{k_{fv}=0,\frac{1}{2},1,\ldots}\Bigr)\,
    \Bigl(\prod_{f\in F}\sym{A}_f\Bigr)\,
    \Bigl(\prod_{e\in E}\sym{A}_e\Bigr)\,
    \Bigl(\prod_{f\in F}\prod_{v\in f_0}\hat K_{k_{fv}}^{(j_f)}\Bigr)\nn\\
  &\times&
    \prod_{v\in V}\Biggl[
    \Bigl(\prod_{\ontop{f\in F\colon}{v\in f_0}}\sum_{n_f,m_f=1}^{2k_{fv}+1}\Bigr)\,
    \Bigl(\prod_{\ontop{e\in E\colon}{v=\del_+(e)}}\sum_{p_e,q_e=1}^{2\ell_e+1}\Bigr)\,
    \Bigl(\prod_{\ontop{e\in E\colon}{v=\del_-(e)}}\sum_{r_e,s_e=1}^{2\ell_e+1}\Bigr)
    P^{(v)}_{\underbrace{\scriptstyle (r_es_e)\ldots}_{\ontop{e\in E\colon}{v=\del_-(e)}}
             \underbrace{\scriptstyle (p_eq_e)\ldots}_{\ontop{e\in E\colon}{v=\del_+(e)}}}\nn\\
  &&\qquad
    \Bigl(\prod_{\ontop{e\in E\colon}{v=\del_+(e)}}
      I^{(+,e)}_{\underbrace{\scriptstyle (n_fm_f)\ldots}_{f\in e_+}(p_eq_e);
                 \underbrace{\scriptstyle (n_fm_f)\ldots}_{f\in e_-}}\Bigr)\,
    \Bigl(\prod_{\ontop{e\in E\colon}{v=\del_-(e)}}
      I^{(-,e)}_{\underbrace{\scriptstyle (n_fm_f)\ldots}_{f\in e_-};
                 \underbrace{\scriptstyle (n_fm_f)\ldots}_{f\in e_+}(r_es_e)}\Bigr)\Biggr].
\end{eqnarray}
This expression looks more complicated than it actually is. We can
explain it in words as follows. There are two types of summations over
representations. These are first the sum over all colourings of the
triangles $f\in F$ with simple representations
$V_{(j_f,j_f^\ast)}=V_{j_f}\otimes V_{j_f}^\ast$ of $\Spin(4)$, and
second the sum over all colourings of the wedges $(f,v)$ with
representations $V_{(k_{fv},k_{fv}^\ast)}$. Here the wedges are
denoted by specifying a dual face $f\in F$ and a four-simplex $v\in
f_0\subseteq V$ whose intersection forms the wedge.

In addition to the face and edge amplitudes $\sym{A}_f$ and
$\sym{A}_e$ already present in the original model~\eqref{eq_zriemann},
there is now an additional amplitude $\hat K_{k_{fv}}^{(j_f)}$ for
each wedge, namely a character expansion coefficient
of~\eqref{eq_character}. For the causal model, this amplitude will in
general be complex.

The amplitude for each four-simplex $v\in V$ is given by the
expression inside the square brackets in~\eqref{eq_backriemann}. It is
given by the usual relativistic $10j$-symbol with a piece of the spin
network $(\ell,P)$ inserted (Figure~\ref{fig_10j}). The various
summations contract the indices of the Barrett--Crane intertwiners
which are denoted by
\begin{eqnarray}
  I^{(+,e)}\colon
    \bigotimes_{f\in e_-} V_{(k_{f\del_+(e)},k_{f\del_+(e)}^\ast)}\to
    \Bigl(\bigotimes_{f\in e_+} V_{(k_{f\del_+(e)},k_{f\del_+(e)}^\ast)}\Bigr)
    \otimes V_{(\ell_e,\ell_e^\ast)},\nn\\
  I^{(-,e)}\colon
    \bigotimes_{f\in e_+} V_{(k_{f\del_-(e)},k_{f\del_-(e)}^\ast)}\to
    \Bigl(\bigotimes_{f\in e_+} V_{(k_{f\del_-(e)},k_{f\del_-(e)}^\ast)}\Bigr)
    \otimes V_{(\ell_e,\ell_e^\ast)},
\end{eqnarray}
using the conventions of~\cite{Pf02a}. Here the sets $e_\pm\subseteq
F$ contain all triangles $f$ that are contained in the boundary of the
tetrahedron $e\in E$ with orientation $\epsilon(e,f)=\pm 1$. If the
intertwiner $P$ of the spin network $(\ell,P)$ is a Barrett--Crane
intertwiner, this amplitude is an evaluated relativistic spin network
and therefore non-negative real. For the Riemannian case, this was
shown in~\cite{Pf02c} whereas for the Lorentzian analogue, this is a
plausible conjecture~\cite{BaCh02}.

Observe that the representations $j_f$ attached to the triangles
appear only in the expressions for the character expansion
coefficients $\hat K_{k_{fv}}^{(j_f)}$. The representations for which
the $10j$-symbols are evaluated, are no longer the $j_f$, but rather
the representations $k_{fv}$ associated with the wedges $(f,v)$.

Equation~\eqref{eq_backriemann} illustrates the impact that the choice
of the causal or Euclidean amplitudes has on the structure and on the
symmetries of the model. The central new feature is the additional
colouring of the wedges with representations. Only for the original
Barrett--Crane model with $\kappa=1$, there exists a significant
simplification because in this case $K_R^{(j)}$ is already an
$\SU(2)$-character. This implies that $\hat K^{(j)}_k=\delta_{jk}$ so
that all wedges $(f,v)$ of a given dual face $f\in F$ are assigned the
same representation $k_{fv}=j_f$. In this special case,
equation~\eqref{eq_backriemann} reduces to the original Barrett--Crane
model~\cite{BaCr98} with a spin network $(\ell,P)$ inserted into its
$10j$-symbols.

Is there a Lorentzian counterpart of the
decomposition~\eqref{eq_character}? In order to derive that formula we
have made use of the identification $S^3\cong\SU(2)$ which does not
have an immediate analogue in the Lorentzian case. Let us reformulate
the argument so that we can generalize it.

The functions $K_{R,X}^{(j)}(x,y)$ all have the symmetry
\begin{equation}
\label{eq_invariance}
  K_{R,X}^{(j)}(gx,gy)=K_{R,X}^{(j)}(x,y),
\end{equation}
for all $x,y\in S^3$ and $g\in\Spin(4)$ acting on $S^3$
(Section~\ref{sect_local}). Due to this symmetry, the function is
already specified if we know its values $f(x):=K_{R,X}^{(j)}(x,e)$
where $e\in S^3$ denotes the north pole. If we write the function
$f\colon S^3\to\C$ as a function on $\Spin(4)$ which is constant on
the left-cosets $gU$ where $U:=\Stab_{\Spin(4)}(e)\cong\SU(2)$ and
$S^3\cong\Spin(4)/U$, the invariance condition~\eqref{eq_invariance}
implies that $f$ is also constant on the right-cosets $Ug$ and
therefore a \emph{zonal spherical function}. These functions are
precisely the characters of $\SU(2)$ using the identification
$S^3\cong\SU(2)$ employed above.

A generalization of~\eqref{eq_character} to the Lorentzian case is now
available since we know that the zonal spherical functions for the
quotient $V\backslash \SL(2;\C)/V$, $V=\Stab_{\SL(2;\C)}(e_t)$,
$e_t=(1,0,0,0)$, are precisely the functions $K_L^{(p)}(x,e_t)$
(see~\eqref{eq_amplorentz}). Therefore we obtain the result that any
$L^2$-function $f\colon H^3_+\times H^3_+\to\C$ which satisfies
\begin{equation}
  f(gx,gy)=f(x,y),
\end{equation}
for all $x,y\in H^3_+$ and $g\in\SL(2;\C)$, is a Plancherel integral
of the form
\begin{equation}
  f(x,y)=\int_0^\infty \hat f(p)\,K_L^{(p)}(x,y)\,p^2\,dp,
\end{equation}
for a suitable function $\hat f\colon R_+\to\C$.

Therefore the strategy which has lead to~\eqref{eq_backriemann}, can
be directly applied to the Lorentzian case. We do not repeat the
analogue of~\eqref{eq_backriemann} here as the required substitutions
are now obvious: replace the sums over half-integers by integrals
$\int_0^\infty p^2\,dp$ and make use of the integral presentation of
the Barrett--Crane intertwiners $I^{(\pm,e)}$. The analogues of the
comments listed below equation~\eqref{eq_backriemann} also apply to
the Lorentzian case.

%------------------------------------------------------------------------------
\subsection{Averaging over the stabilizer}
%------------------------------------------------------------------------------

In~\cite{Pf02a} we have developed the quantum geometry of the
Barrett--Crane model in the connection picture. This includes in
particular the interpretation of the integrals over $S^3$ or $H^3_+$
that appear in the Barrett--Crane intertwiner, as integrals over
possible directions of the vectors normal to the tetrahedra. The fact
that there are two such variables for each tetrahedron was interpreted
as the consequence of a non-trivial parallel transport which is
associated with each tetrahedron and which maps the first normal to
the second one. This parallel transport, however, is not a full
$\Spin(4)$- or $\SL(2;\C)$-parallel transport, but it is rather
specified only up to elements of the stabilizer which leave the first
normal vector fixed. It was then possible to express what is the
difference between $BF$-theory and the Barrett--Crane model. It is
precisely this averaging over the stabilizers. For more details,
see~\cite{Pf02a}.

The geometric interpretation was developed originally for Riemannian
signature, see Section~4.2 of~\cite{Pf02a}. These results can be
easily translated to Lorentzian signature by the replacement of
$\Spin(4)/\SU(2)\cong S^3$ by $\SL(2,\C)/\SU(2)\cong H^3_+$. The
comparison of the Barrett--Crane model with $BF$-theory in Section~4.3
of~\cite{Pf02a} relies on Lemma~4.4 therein whose generalization we
sketch in the following.

Let $t^{(n,p)}_{(j_1m_1)(j_2m_2)}$ denote a representative function of
$\SL(2;\C)$ in a representation of type $V_{(n,p)}$, $n\in\Z$, $p>0$,
of the principal series. We realize the representation $V_{(n,p)}$ as
a suitable space of sections of a line bundle $S^3\to S^2$ and obtain
an orthonormal basis from the spherical functions on the total space
$S^3$ so that the range of the indices of the representative functions
is given by $j_\ell=|\frac{n}{2}|, |\frac{n}{2}|+1,\ldots$ and
$m_\ell\in\{-j_\ell,-j_\ell+1,\ldots,j_\ell\}$ for $\ell=1,2$. It is
then possible to show that
\begin{equation}
  \int_{\SU(2)}t^{(n,p)}_{(j_1m_1)(j_2m_2)}(gu)\,du=\left\{
    \begin{matrix}
      0,&\mbox{if}\quad n\neq 0,\\
      H^{(p)}_{j_1m_1}\delta_{j_20}\delta_{m_20},&\mbox{if}\quad n=0.
    \end{matrix}\right.
\end{equation}
Here $\SU(2)\subseteq\SL(2,\C)$ is embedded as the stabilizer of
$e_t=(1,0,0,0)\in H^3_+$, and the functions $H^{(p)}_{jm}$ form an
orthonormal basis of functions $H^3_+\to\C$,
see~\eqref{eq_gelfand}. This shows that there exists an
$\SU(2)$-invariant subspace of $V_{(n,p)}$ only if the representation
is simple, $n=0$, and that this subspace is one-dimensional. We
therefore obtain basis functions on $H^3_+\cong\SL(2;\C)/SU(2)$ from
representative functions of $\SL(2;\C)$ by averaging over the right
$\SU(2)$-action.

Now we consider two points $x,y\in H^3_+$. For each $z\in H^3_+$,
there exists some boost $b_z\in\SL(2;\C)$ such that $b_z(e_t)=z$. The
group elements $g\in\SL(2;\C)$ that map $gy=x$ are of the form
\begin{equation}
  g=\alignidx{b_xb_y^{-1}u_y},
\end{equation}
where $u_y\in\Stab_{\SL(2;\C)}(y)\cong\SU(2)$. However, if
$u_y\in\Stab(y)$, then $\alignidx{b_y^{-1}u_yb_y}\in\Stab(e_t)$ and
conversely, therefore
\begin{equation}
 g=\alignidx{b_xu_tb_y^{-1}},
\end{equation}
for some $u_t\in\Stab(e_t)$. We use this parametrization of $g$ and
average over the stabilizer,
\begin{equation}
  \int_{\SU(2)}t^{(n,p)}_{(j_1m_1)(j_2m_2)}(\alignidx{b_xub_y^{-1}})\,du
  =\delta_{n0}H^{(p)}_{j_1m_1}(b_x)\overline{H^{(p)}_{j_2m_2}(b_y)}.
\end{equation}
The construction of~\cite{Pf02a} then says that for any holonomy
$g\in\SL(2,\C)$ at an edge, it matters only how $g$ acts on
$H^3_+$. Therefore we choose some $x\in H^3_+$, calculate $y=gx$ and
average over the stabilizer ambiguity. The integration over the group
which is present in the path integral then results in the desired
integrations over $H^3_+$. This is the Lorentzian analogue of
Lemma~4.4 of~\cite{Pf02a}.

% ==============================================================================
%
\section{Discussion}
%
% ==============================================================================
\label{sect_conclusion}

What we have explained in Sections~\ref{sect_wick}
and~\ref{sect_observe}, the Euclidean reconstruction and the analysis
of the degrees of freedom of the Barrett--Crane model from an
understanding of its local symmetries, is only one motivation for
studying the observables in the connection picture. Another motivation
arises from the observation~\cite{Pf02a} that for some edge
amplitudes, the partition function in the connection picture is
particularly simple and resembles a spin model (just `spin', not `spin
foam') with variables in $S^3$ or $H^3_+$ with local interaction terms
at the faces. In particular, no evaluations of $10j$-symbols are
necessary in this case which makes numerical simulations
computationally cheaper. One has just to tabulate the interaction
terms.

For the original Riemannian model it was observed in both
formulations, in the connection picture~\cite{Pf02a} and in the
representation picture~\cite{BaCh02c}, that the dominant
configurations of the partition function often correspond to
degenerate geometries. With the results presented here, there are two
new developments which can modify this conclusion. This is first the
introduction of the constant $\kappa$ (Section~\ref{sect_constant})
which provides a natural way of controlling the width of the peaks in
the picture of~\cite{Pf02a}. This is what a coupling constant
[temperature] in a path integral [Statistical Mechanics] model
typically does. The constant $\kappa$ may play an important role when
one tries to locate a critical point at which one can renormalize the
model. Second, the causal and the Euclidean model have amplitudes very
different from the original model. In the connection picture, the
Euclidean version can be studied by exactly the same techniques as the
original version so that one can start to investigate and compare the
models and their physical interpretation. The transformation of
Section~\ref{sect_transback} then allows us to perform the same
studies in the representation picture.

Finally, we note that all our formulas for partition functions, matrix
elements and expectation values use the language of generic
two-complexes. These are not restricted to be dual to a given
triangulation. The only exceptions were the motivating steps which
explicitly involved results from Regge calculus which are available
only on triangulations.

%------------------------------------------------------------------------------
\acknowledgments
%------------------------------------------------------------------------------

The author would like to thank Etera Livine and Daniele Oriti for
discussions.

\end{document}